\documentstyle[epsfig]{cjpsuppl}
\newcommand{\newc}{\newcommand}
\newc{\ra}{\rightarrow}
\newc{\lra}{\leftrightarrow}
\newc{\beq}{\begin{equation}}
\newc{\eeq}{\end{equation}}
\newc{\barr}{\begin{eqnarray}}
\newc{\earr}{\end{eqnarray}}

\newcommand{\lsim}   {\mathrel{\mathop{\kern 0pt \rlap
  {\raise.2ex\hbox{$<$}}}
  \lower.9ex\hbox{\kern-.190em $\sim$}}}
\newcommand{\gsim}   {\mathrel{\mathop{\kern 0pt \rlap
  {\raise.2ex\hbox{$>$}}}
  \lower.9ex\hbox{\kern-.190em $\sim$}}}
\begin{document}
\title {Searching for Dark
Matter-\\Can it be directly detected?}
\authori{J.D.\, Vergados}
\addressi{Physics Department, University of Ioannina, Ioannina, Greece
\\and\\Physics Department, University of Cyprus, Nicosia, Greece}
\authorii{}    \addressii{}
\authoriii{}   \addressiii{}
\authoriv{}    \addressiv{}
\authorv{}     \addressv{}
\authorvi{}    \addressvi{}
\headtitle{Searching for Dark Matter \ldots} \headauthor{J.D.
Vergados} \lastevenhead{J.D. Vergados: Searching for Dark
Matter-Can it be directly detected? \ldots} 
\refnum{}
\daterec{26 January 2003;\\final version January 26 2004}
\suppl{A}  \year{2004} \setcounter{page}{21}
\maketitle
\begin{abstract}
In this paper we review the theoretical issues involved in the
direct detection of supersymmetric (SUSY) dark matter. After a
brief discussion of the allowed SYSY parameter space we focus on
the determination of the traditional neutralino detection rates,
in experiments which measure the energy of the recoiling nucleus,
such as the coherent and spin induced rates and the dependence of
the rate on the motion of the Earth (modulation effect). Then we
examine the novel features  appearing in directional experiments,
which detect the recoiling nucleus in a given direction. Next we
estimate the branching ratios for transitions to accessible
excited nuclear levels. Finally we comment on the event rates
leading to the atom ionization and subsequent detection of the
outgoing electrons.
  \end{abstract}
\section{Introduction}
The combined MAXIMA-1 \cite{MAXIMA-1}, BOOMERANG \cite{BOOMERANG},
 DASI \cite{DASI} and COBE/DMR Cosmic Microwave Background (CMB)
observations \cite{COBE} as well as the recent WMAP data
\cite{SPERGEL} imply that the Universe is flat \cite{flat01}and
that most of the matter in the Universe is Dark, i.e. exotic.
Crudely speaking and easy to remember one has:
$$\Omega_b=0.05, \Omega _{CDM}= 0.30, \Omega_{\Lambda}= 0.65$$
for the baryonic, dark matter and dark energy fractions
respectively.

  These observations, however, do not tell us anything about the
  particle nature of dark matter. This can only be accomplished
  through direct observation. Many experiments are currently under
 way aiming at this goal.

 Supersymmetry naturally provides candidates for the dark matter constituents
\cite{Jung},\cite{GOODWIT}-\cite{ELLROSZ}.
 In the most favored scenario of supersymmetry the
lightest supersymmetric particle (LSP) can be simply described as
a Majorana fermion, a linear combination of the neutral components
of the gauginos and higgsinos
\cite{Jung},\cite{GOODWIT}-\cite{ref2}.

In all calculations performed so far, the obtained event rates are
 quite low and perhaps unobservable in the near future. So one
has to search for characteristic signatures associated with this
reaction. Such are the modulation of the event rates with the
motion of the Earth (modulation effect) and the correlation of the
observed rates of directionally sensitive experiments with the
motion of the sun \cite{JDV03,Verg}. Transitions to low energy
excited nuclear states have also been considered \cite{VQS03}.
Quite recently it has been found that the detection of electrons,
following the collision of the neutralino with the nucleus may
offer another option \cite{JDV04} to be exploited by the
experiments.

\section{The Essential Theoretical Ingredients  of Direct Detection.}
 The possibility of dark matter detection hinges on the nature of its
 constituents. Here we will assume that such a constituent is the
 lightest supersymmetric particle or LSP.
  Since this particle is expected to be very massive, $m_{\chi} \geq 30 GeV$, and
extremely non relativistic with average kinetic energy $T \approx
50~keV (m_{\chi}/ 100 GeV)$, it can be directly detected
~\cite{Jung}-\cite{KVprd} mainly via the recoiling of a nucleus
(A,Z) in elastic scattering. In this paper, however, we will
consider alternative possibilities.

 The event rate for all such  processes can be computed from the
 following ingredients:
\begin{enumerate}
\item An effective Lagrangian at the elementary particle (quark)
level obtained in the framework of supersymmetry as described ,
e.g., in Refs~\cite{ref2,JDV96}. \item A well defined procedure
for transforming the amplitude obtained using the previous
effective Lagrangian from the quark to the nucleon level, i.e. a
quark model for the nucleon. This step is not trivial, since the
obtained results depend crucially on the content of the nucleon in
quarks other than u and d.
 \item Nuclear matrix elements
\cite{Ress}$-$\cite{DIVA00},\cite{VQS03}, \cite{JDV04},
 obtained with as reliable as possible
many body nuclear wave functions. Fortunately in the most studied
case of the scalar coupling the situation is quite simple, since
then one needs only the nuclear form factor. Some progress has
also been made in obtaining reliable static spin matrix elements
and spin response functions \cite{DIVA00},\cite{VQS03}.
\end{enumerate}

 The calculation of this cross section  has become pretty standard.
 One starts with
representative input in the restricted SUSY parameter space as
described in the literature for the scalar interaction~\cite{Gomez,ref2}
 (see also Arnowitt
and Dutta \cite{ARNDU}).

  It is worth exploiting the contribution of the axial current in the
 direct neutralino detection, since, in addition, it may populate
 excited  nuclear states, if they happen to be so low in energy that they become
 accessible to  the low energy neutralinos \cite{VQS03}. Models which can
 lead to detectable spin cross sections have recently been
 proposed \cite{CCN03} \cite{CHATTO} \cite{WELLS} \cite{LIN01}.

 Once the LSP-nucleon cross section is known,
the LSP-nucleus cross section can be obtained. The differential
cross section with respect to the energy transfer $Q$ for a given
LSP velocity $\upsilon$ can be cast in the form
\begin{equation}
d\sigma (u,\upsilon)= \frac{du}{2 (\mu _r b\upsilon )^2}
[(\bar{\Sigma} _{S}F^2(u)
                       +\bar{\Sigma} _{spin} F_{11}(u)]
\label{2.9}
\end{equation}
where we have used a dimensionless variable $u$, proportional to
$Q$, which is found convenient for handling the nuclear form
factors \cite{KVprd} $F(u)~,~F_{11}(u)$, namely
$u=\frac{Q}{Q_0}~~,~~Q_{0}\approx 40 \times A^{-4/3}~MeV.$ $\mu_r$
is the reduced LSP-nucleus mass and $b$ is (the harmonic
oscillator) nuclear size parameter. Furthermore
\begin{equation}
\bar{\Sigma} _{S} = \sigma^S_{p,\chi^0} A^2 \mu^2_r,
  \bar{\Sigma}
_{spin}  =  \mu^2_r
                           \sigma^{spin}_{p,\chi^0}~\zeta_{spin},
\zeta_{spin}= \frac{1}{3(1+\frac{f^0_A}{f^1_A})^2}S(u)
 \label{2.10}
\end{equation}
$\sigma^{spin}_{p,\chi^0}$ and $\sigma^{s}_{p,\chi^0}$ are the nucleon
 cross-sections associated with the spin and the scalar interactions
respectively and
 \beq
S(u)=[(\frac{f^0_A}{f^1_A} \Omega_0(0))^2
\frac{F_{00}(u)}{F_{11}(u)}
  +  2\frac{f^0_A}{ f^1_A} \Omega_0(0) \Omega_1(0)
\frac{F_{01}(u)}{F_{11}(u)}+  \Omega_1(0))^2  \, ]
\label{S(u)}
\eeq
 The definition of the spin response functions $F_{ij}$,
  with $i,j=0,1$ isospin indices, can be found
 elsewhere \cite{DIVA00}.

Some  static spin matrix elements \cite{DIVA00}, \cite{Ress}, \cite{KVprd}
for some nuclei of interest are given in
table \ref{table.spin}
\begin{table}
\caption{ The static spin matrix elements for various nuclei. For
light nuclei the calculations are from Divari et al (see text) .
For $^{127}I$ the results are from Ressel and Dean (see text) (*)
and  the
 Jyvaskyla-Ioannina collaboration (private communication)(**).
 For $^{209}Pb$ they were obtained previously (see text).
} \label{table.spin}
\begin{center}
\begin{tabular}{lrrrrrr}
 &   &  &  &  &   & \\
 & $^{19}$F & $^{29}$Si & $^{23}$Na  & $^{127}I^*$ & $ ^{127}I^{**}$ & $^{207}Pb^+$\\
\hline
    &   &  &  &  &    \\
$[\Omega_{0}(0)]^2$         & 2.610   & 0.207  & 0.477  & 3.293   &1.488 & 0.305\\
$[\Omega_{1}(0)]^2$         & 2.807   & 0.219  & 0.346  & 1.220   &1.513 & 0.231\\
$\Omega_{0}(0)\Omega_{1}(0)$& 2.707   &-0.213  & 0.406  &2.008    &1.501&-0.266\\
$\mu_{th} $& 2.91   &-0.50  & 2.22  &    &\\
$\mu_{exp}$& 2.62   &-0.56  & 2.22  &    &\\
$\frac{\mu_{th}(spin)}{ \mu_{exp}}$& 0.91   &0.99  & 0.57  &    &  &\\
\end{tabular}
\end{center}
\end{table}
\section{Rates}
The differential non directional  rate can be written as:
\begin{equation}
dR_{undir} = \frac{\rho (0)}{m_{\chi}} \frac{m}{A m_N}
 d\sigma (u,\upsilon) | {\boldmath \upsilon}|
\label{2.18}
\end{equation}
 Where   $\rho (0) = 0.3 GeV/cm^3$ is the LSP density in our vicinity,
 m is the detector mass, $m_{\chi}$ is the LSP mass and
$d\sigma(u,\upsilon )$ was given above.\\
 The directional differential rate, in the direction $\hat{e}$ of the
 recoiling nucleus, is given by :
\begin{eqnarray}
dR_{dir} &=& \frac{\rho (0)}{m_{\chi}} \frac{m}{A m_N}
|\upsilon| \hat{\upsilon}.\hat{e} ~\Theta(\hat{\upsilon}.\hat{e})
 ~\frac{1}{2 \pi}~
d\sigma (u,\upsilon)\\
\nonumber & &\delta(\frac{\sqrt{u}}{\mu_r \upsilon
b\sqrt{2}}-\hat{\upsilon}.\hat{e})
 ~~,~ \Theta (x)= \left \{
\begin{array}{c}1~,x>0\\0~,x<0 \end{array} \right \}
 \label{2.20}
\end{eqnarray}

The LSP is characterized by a velocity distribution. For a given
velocity distribution f(\mbox{\boldmath $\upsilon$}$^{\prime}$),
 with respect to the center of the galaxy,
One can find the  velocity distribution in the lab frame
$f(\mbox{\boldmath $\upsilon$},\mbox{\boldmath $\upsilon$}_E)$ by
writing \hspace{1.0cm}\mbox{\boldmath $\upsilon$}$^{'}$=
          \mbox{\boldmath $\upsilon$}$ \, + \,$ \mbox{\boldmath $\upsilon$}$_E
 \,$ ,
\hspace{1.0cm}\mbox{\boldmath $\upsilon$}$_E$=\mbox{\boldmath
$\upsilon$}$_0$+
 \mbox{\boldmath $\upsilon$}$_1$.
\mbox{\boldmath $\upsilon$}$_0 \,$  is the sun's velocity (around
the center of the galaxy), which coincides with the parameter of
the Maxwellian distribution, and \mbox{\boldmath $\upsilon$}$_1
\,$ the Earth's velocity
 (around the sun).
Thus the above expressions for the rates must be folded with the
LSP velocity
 distribution. We will distinguish two possibilities:
\begin{enumerate}
 \item The direction of the recoiling nucleus is not observed.\\
 The non-directional differential rate is now given by:
\begin{equation}
\Big<\frac{dR_{undir}}{du}\Big> = \Big<\frac{dR}{du}\Big> =
\frac{\rho (0)}{m_{\chi}} \frac{m}{Am_N} \sqrt{\langle
\upsilon^2\rangle } {\langle \frac{d\Sigma}{du}\rangle }
\label{3.11}
\end{equation}
where
\begin{equation}
\langle \frac{d\Sigma}{du}\rangle =\int
           \frac{   |{\boldmath \upsilon}|}
{\sqrt{ \langle \upsilon^2 \rangle}}
 f(\mbox{\boldmath $\upsilon$},\mbox{\boldmath $\upsilon$}_E)
                       \frac{d\sigma (u,\upsilon )}{du} d^3
 \mbox{\boldmath $\upsilon$}
\label{3.12a}
\end{equation}
\item  The direction $\hat{e}$ of the recoiling nucleus is observed.\\
In this case the directional differential rate is given as above
with:
\begin{eqnarray}
\langle (\frac{d\Sigma}{du})_{dir}\rangle &=&\int \frac{
\mbox{\boldmath $\upsilon$}.\hat{e}~
            \Theta( \mbox{\boldmath $\upsilon$}.\hat{e})}
{\sqrt{ \langle \upsilon^2 \rangle}}
 f(\mbox{\boldmath $\upsilon$},\mbox{\boldmath $\upsilon$}_E)
                       \frac{d\sigma (u,\upsilon )}{du}\\
\nonumber & &\frac{1}{2 \pi}
 \delta(\frac{\sqrt{u}}{\mu_r b \upsilon}-\hat{\upsilon}.\hat{e}) d^3
 \mbox{\boldmath $\upsilon$}
\label{3.12j}
\end{eqnarray}
\end{enumerate}
To obtain the total rates one must integrate  the two previous
expressions
  over the energy transfer from
$Q_{min}$ determined by the detector energy cutoff to $Q_{max}$
determined by the maximum LSP velocity (escape velocity, put in by
hand in the Maxwellian distribution), i.e.
$\upsilon_{esc}=2.84~\upsilon_0$, $\upsilon_0=229~Km/s$.
\section{Results}
We will specialize  the above results in the following cases:
\subsection{Non directional unmodulated rates}
Ignoring the motion of the Earth the total non directional rate is
given by
\begin{equation}
R =  \bar{R}\, t(a,Q_{min}) \,
\label{3.55f}
\end{equation}
$$ \bar{R}=\frac{\rho (0)}{m_{\chi^0}} \frac{m}{Am_p}~
              (\frac{\mu_r}{\mu_r(p)})^2~ \sqrt{\langle
v^2 \rangle } [\sigma_{p,\chi^0}^{S}~A^2+
 \sigma _{p,\chi^0}^{spin}~\zeta_{spin}]$$
where $t$ is the modification of the total rate due to the
 folding and nuclear structure effects.
 $t$  depends on
$Q_{min}$, i.e.  the  energy transfer cutoff imposed by the detector
 and the parameter $a$ introduced above.
 All SUSY parameters, except the LSP mass, have been absorbed in $\bar{R}$.

Via  Eq. (\ref{3.55f}) we can, if we wish,   extract the nucleon cross
 section from the data.
For most of the allowed parameter space the obtained results are undetectable.
As it has already been mentioned it is possible to obtain detectable rates in
 the case of the coherent mode. Such results have, e.g. been obtained
by Cerdeno {\it et al} \cite{CERDENO} with non universal set of parameters
 and the Florida group \cite{BAER03}.

\begin{figure}
\begin{center}
\includegraphics[height=.15\textheight]{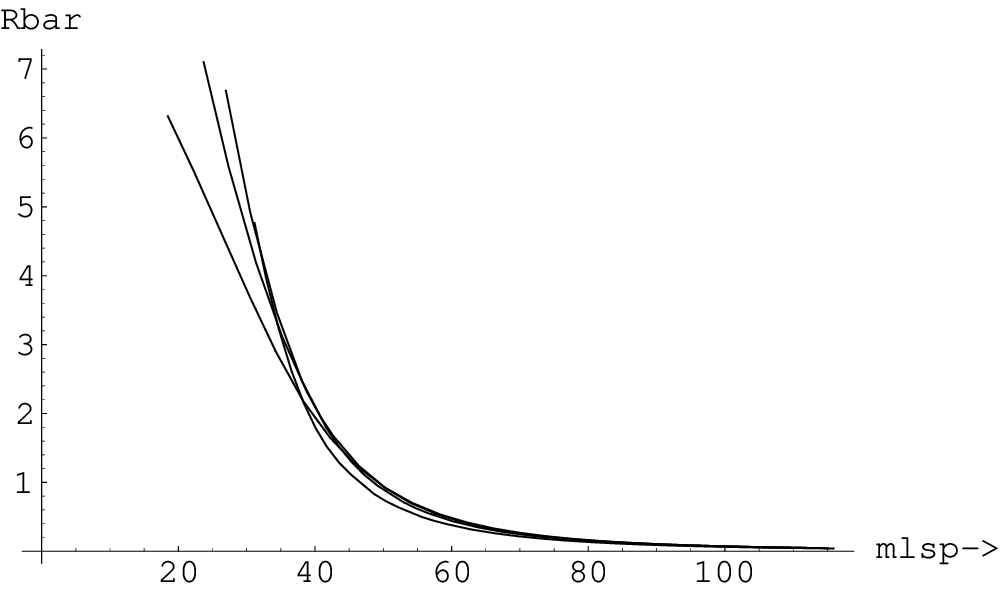}
\includegraphics[height=.15\textheight]{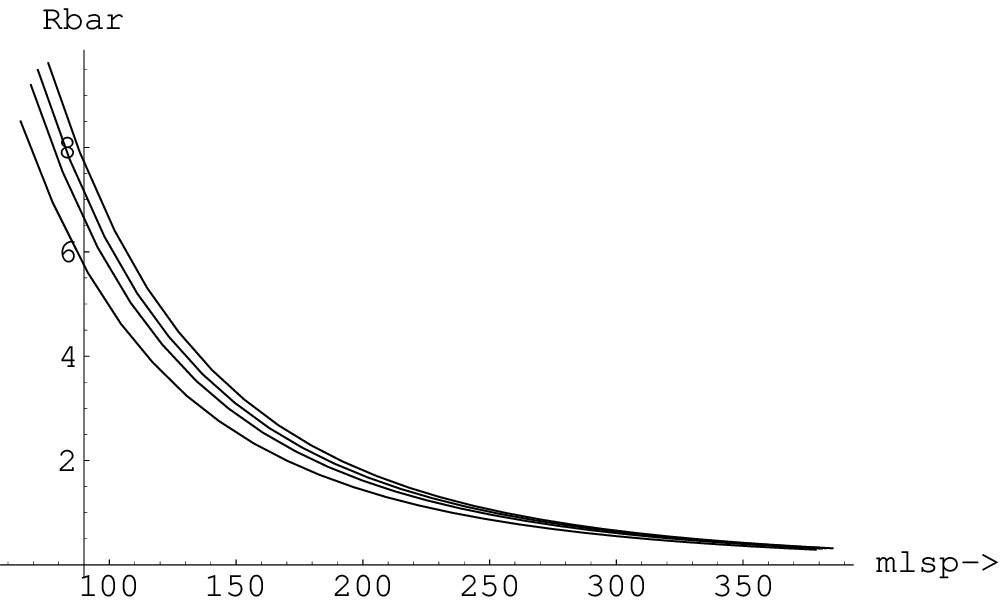}\\
\includegraphics[height=.15\textheight]{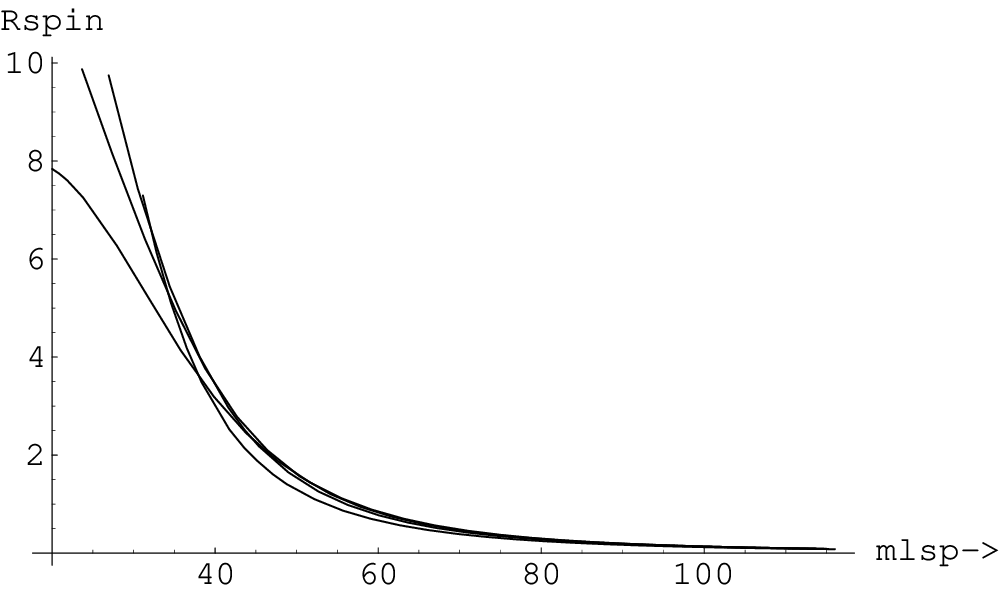}
\includegraphics[height=.15\textheight]{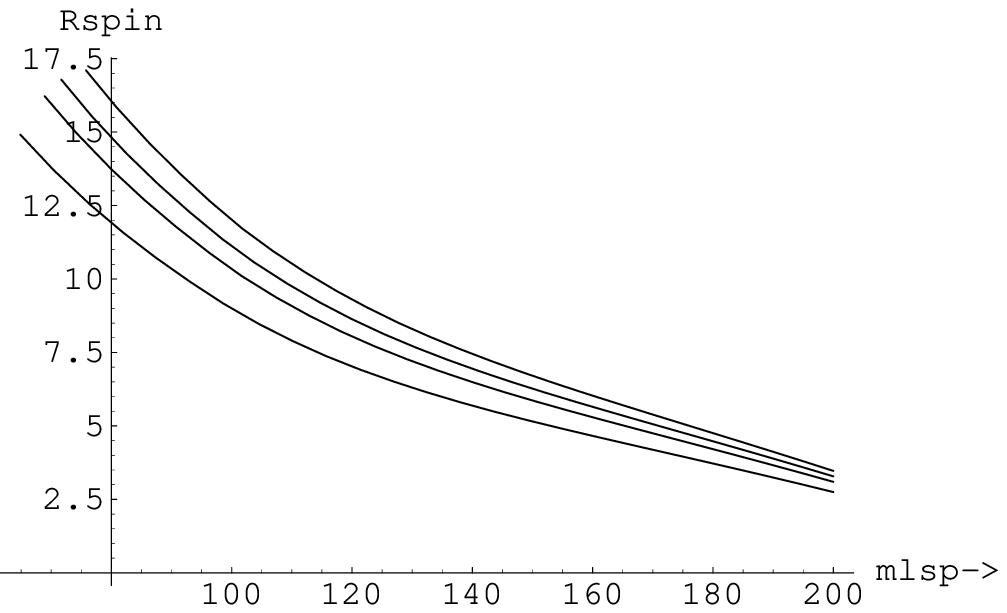}
\caption{
 On top:The quantity $\bar{R},~\approx$ event rate for $Q_{min}=0$,
 associated with  the spin
contribution in the case of the $A=19$ system (for the definition
of the parameters see text). bottom: The event rate, associated
with the spin contribution in the case of the $A=127$ system (for
notation see our earlier work \cite{VQS03}).
 \label{draw1} }
\end{center}
\end{figure}

For the target $^{19}F$ are shown in Fig. \ref{draw1} (top), while
for $^{127}I$ the corresponding results are shown in Fig.
\ref{draw1}.
\subsection{Modulated Rates}
If the effects of the motion of the Earth around the sun are included, the total
 non directional rate is given by
\begin{equation}
R =  \bar{R}\, t \, [(1 +  h(a,Q_{min})cos{\alpha})]
\label{3.55j}
\end{equation}
with  $h$  the modulation amplitude and $\alpha$ is the phase of the Earth, which is
zero around June 2nd. The modulation amplitude would be an excellent signal in
discriminating against background, but unfortunately it is very small, less
 than two per cent (see table \ref{table1.gaus}).
 \begin{table}
\caption{ The parameters $t$, $h$, $\kappa,h_m$ and $\alpha_m$ for
the
 isotropic Gaussian
 velocity distribution and $Q_{min}=0$. The results presented are  associated
 with the spin contribution, but those
 for the coherent mode are similar. The results shown are for the light
systems. For intermediate and heavy nuclei there is a dependence
on the LSP mass. $+x$ is
 radially  out of the galaxy ($\Theta=\pi/2,\Phi=0$), $+z$ is in the sun's
 direction of motion ($\Theta=0$) and
$+y$ is vertical to the plane of the galaxy
($\Theta=\pi/2,\Phi=\pi/2$) so that
 $(x,y,x)$ is right-handed. $\alpha_m=0,1/2,1,3/2$ means
that the maximum occurs on the 2nd of June, September, December
and March
 respectively.
\label{table1.gaus}}
\begin{center}
\begin{tabular}{lrrrrrr}
& & & & & &      \\
type&t&h&dir &$\kappa$ &$h_m$ &$\alpha_m$ \\
\hline
& & & & & &      \\
& &&+z        &0.0068& 0.227& 1\\
dir& & &+(-)x      &0.080& 0.272& 3/2(1)\\
& & &+(-)y        &0.080& 0.210& 0 (1)\\
& & &-z         &0.395& 0.060& 0\\
\hline
all&1.00& & && & \\
all& & 0.02& & & & \\
\hline
\end{tabular}
\end{center}
\end{table}
Furthermore for intermediate and heavy nuclei, it can even change
sign
for sufficiently  heavy LSP. So in our opinion a better signature is provided
 by directional experiments, which measure the direction of the recoiling nucleus.
\subsection{Directional Rates.}
Since the sun is moving around the galaxy in a directional
experiment, i.e. one in which the direction of the recoiling
nucleus is observed, one expects a strong correlation of the event
rate with the motion of the sun. The directional rate can be
written as:
 \barr
 R_{dir}  &=&   \frac{t_{dir}} {2 \pi} \bar{R}
  [1 + h_m  cos {(\alpha-\alpha_m~\pi)}]\\
 \nonumber
 &=& \frac{\kappa} {2 \pi} \bar{R}~t
            [1 + h_m  cos {(\alpha-\alpha_m~\pi)}]
\label{4.56b} \earr
  where $h_m$ is the modulation,
 $\alpha_m $ is the shift in the phase of the Earth $\alpha$
and $\kappa/(2 \pi)$ is the reduction factor
of the unmodulated directional rate relative
to the non-directional one.
The parameters  $\kappa~,~h_m~,~\alpha_m$ depend on the direction of
 observation:
 The above parameters are shown in Table \ref{table1.gaus}
 parameter $t_{dir}$  for a typical LSP mass $100~GeV$ is shown in
for the targets $A=19$. For heavier targets the depend on the LSP
mass \cite{JDV03}.


\subsection{ Rates to excited states}
 Transitions to excited states are possible only for nuclear systems
 characterized by excited states at sufficiently low energies with quantum
 numbers, which allow for Gamow-Teller transitions. One such system is
 $^{127}I$, which, fortunately, can serve as a target for the recoil
 experiment.

 This nucleus has a ground state $5/2^+$ and a first excited state
a $7/2^+$ at $57.6 keV$. As it has already been mentioned  it is a
popular target for dark matter detection.  As a result the
structure of its ground state has been studied theoretically by a
lot of groups (for references see \cite{VQS03}). We find
$\Omega^2_0=\Omega^2_1=\Omega_0 \Omega_1=0.164,~0.312$ for the
ground state and the excited state respectively.

\begin{figure}
\begin{center}
\includegraphics[height=.10\textheight]{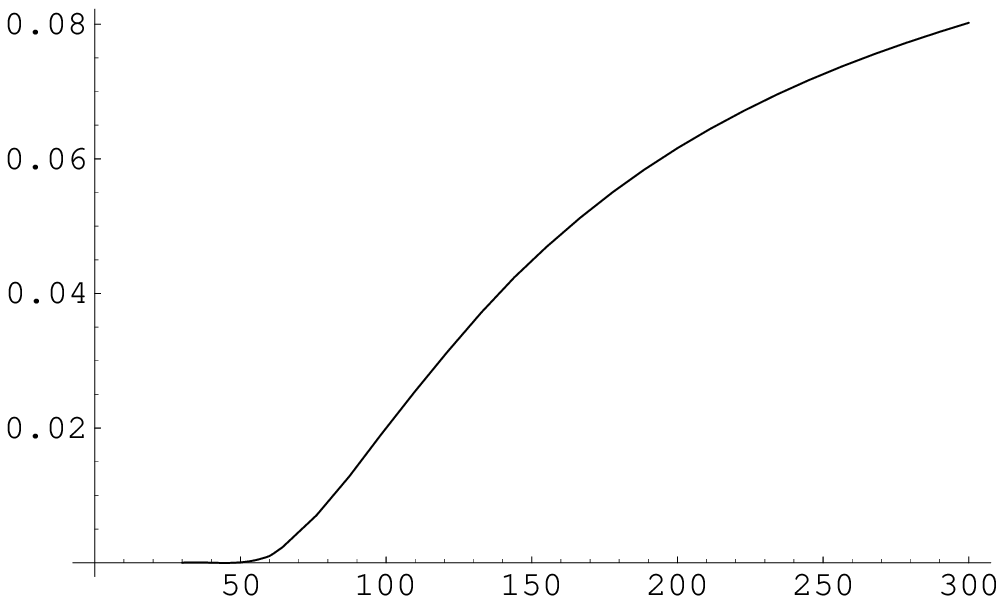}
\includegraphics[height=.10\textheight]{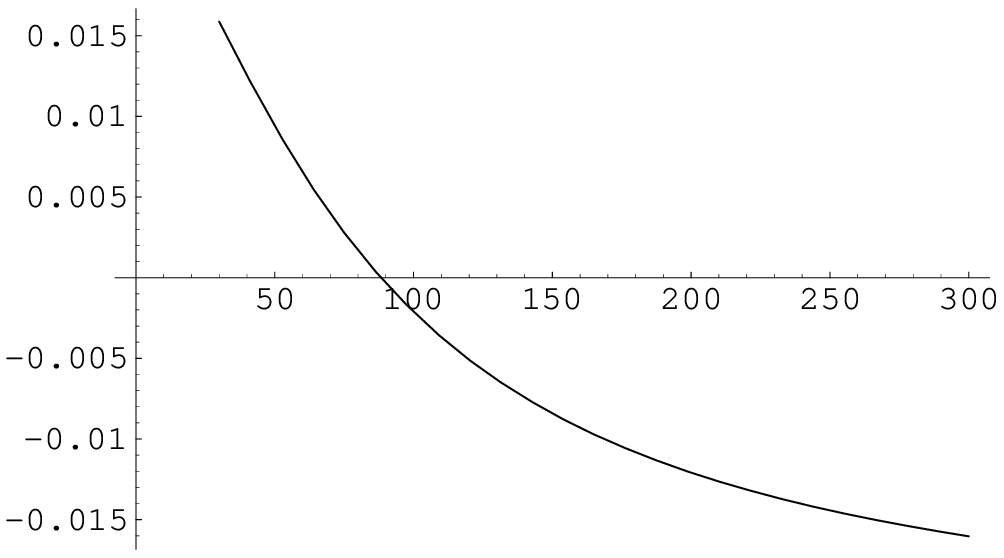}
\includegraphics[height=.10\textheight]{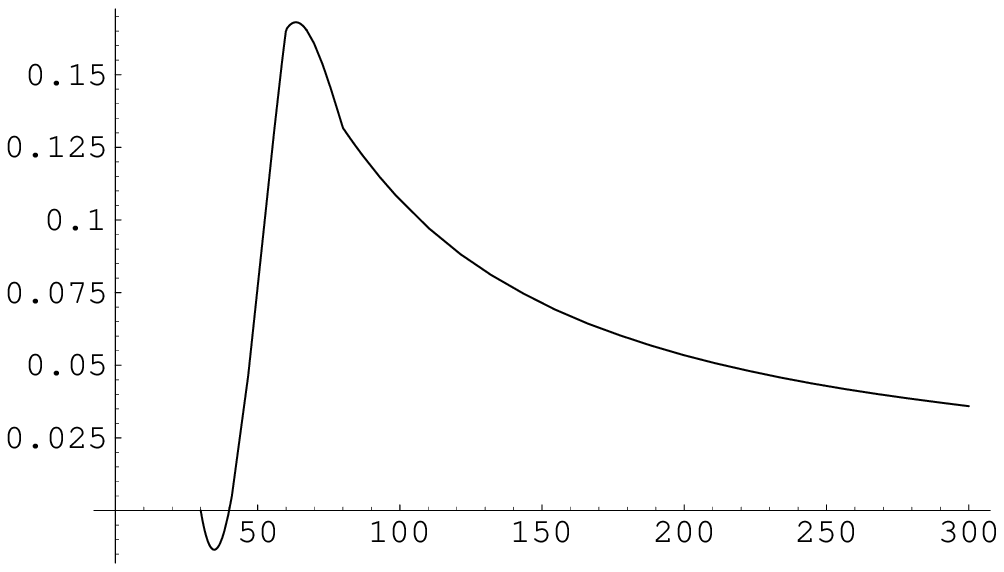}
\caption{ On the left we show he ratio of the rate to the excited
state divided by that of the ground state for $^{127}I$ assuming
that the static spin matrix element of the transition from the
ground to the excited state is a factor of 1.902 larger than that
involving the ground state, but  the spin response functions are
the same. Next to it we show the modulation amplitudes for the
ground and the excited states respectively. The results were
obtained for no threshold cut off  ($Q_{min}=0$).
  \label{exc1} }
\end{center}
\end{figure}

\begin{figure}
\begin{center}
\includegraphics[height=.10\textheight]{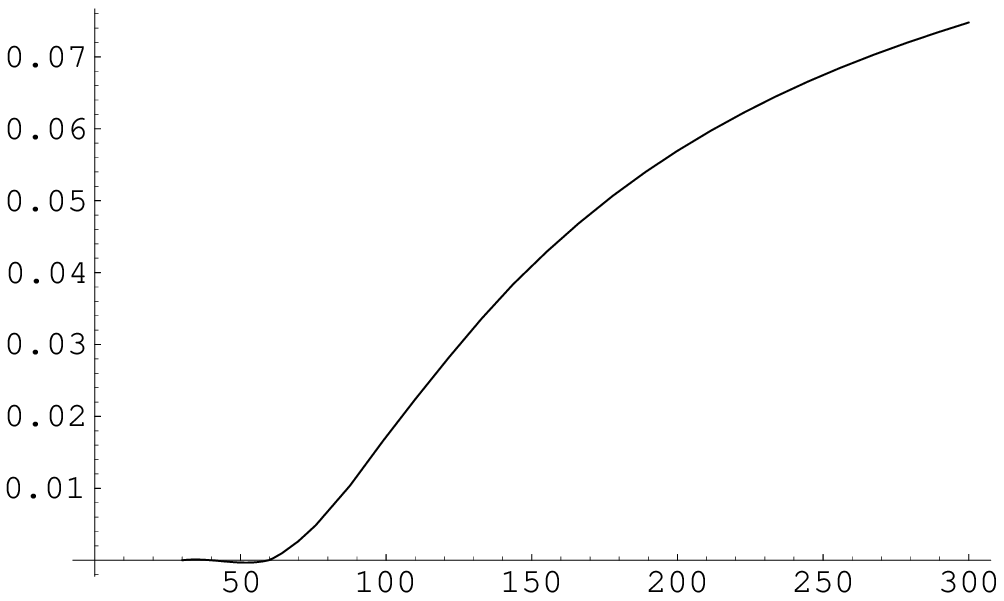}
\includegraphics[height=.10\textheight]{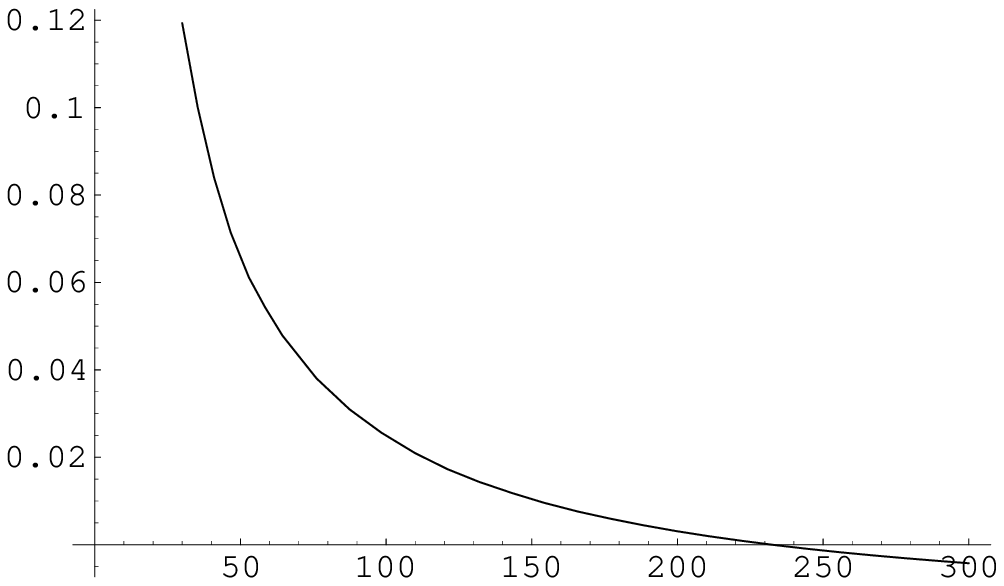}
\includegraphics[height=.10\textheight]{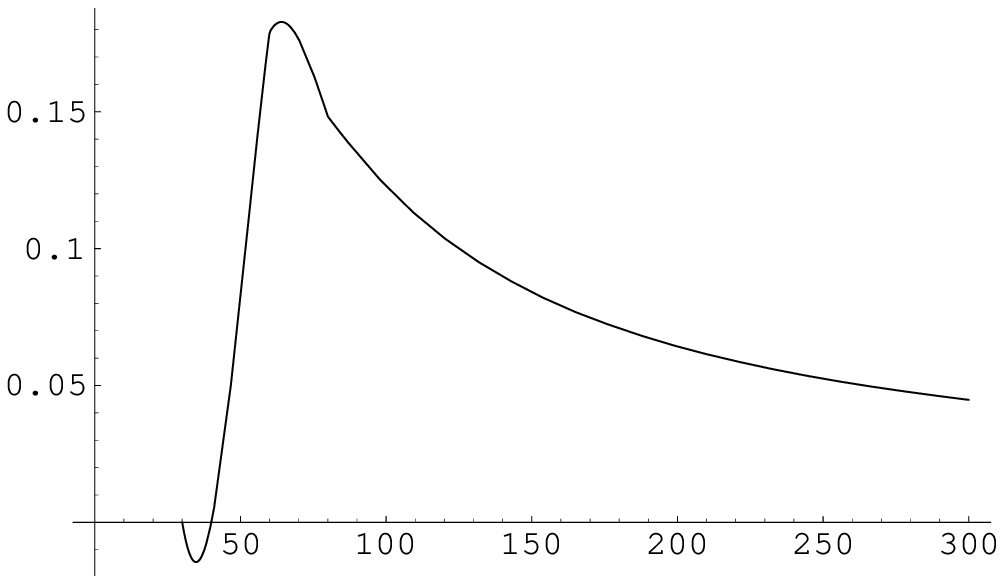}
\caption{ The the same as in Fig. \ref{exc1} for a lower energy
cutoff of $Q_{min}=10 keV$.
 \label{exc2} }
\end{center}
\end{figure}

  In presenting our results it is advantageous to compute the branching
ratio. In addition to factoring out most of the uncertainties
connected with the SUSY parameters and the structure of the
nucleon, we expect the ratio of the two spin matrix elements to be
more reliable than their absolute values. Taking the ratio of the
static spin matrix elements to be $1.90$ and assuming that the
spin response functions are identical, we calculated the branching
ratio , which is exhibited in  Figs \ref{exc1} and \ref{exc2}. We
notice that the dependence on $Q_{min}$ is quite mild.
 From Figs \ref{exc1} and \ref{exc2} we notice that the relative modulation is
 higher when the phase space is restricted by $Q_{min}$ and $E_{exc}$ at the
expense, of course, of the total number of counts.
\subsection{Detection of ionization electrons}

 The experiments may also attempt to detect the ionization electrons,
 which maybe emitted during the LSP-nucleus collision \cite{JDV04}. This
 interesting signature can occur, both for the coherent and the spin
 mode,
 with a branching ratio between $2\%$ and $8\%$. Due to lack of space,
 however, we are not going to elaborate further on this point.

\section{Conclusions}
In the present paper we have discussed the parameters, which
describe the event rates for direct detection of SUSY dark matter.
In the coherent case, only in a small segment of the allowed
 parameter space the rates are above
 the present experimental goals ~\cite{Gomez,ref2,ARNDU}, which, of
 course,
may  be improved  by two or three orders of magnitude in the
planned experiments \cite{CDMS}-\cite{GENIUS}. In the case of the
spin contribution only in models with large higgsino components of
the LSP one can obtain rates, which may be detectable, but in this
case, except in special models, the bound on the relic LSP
abundance may be  violated. Thus in both cases the expected rates
 are small.
  Thus one feels compelled to look for
characteristic experimental signatures for background reduction,
such as correlation of the event rates with the motion of the
Earth (modulation effect) and angular correlation of the
directional rates with the direction of motion  of the sun on top
of their seasonal modulation. Such experiments are currently under
way, like the UKDMC DRIFT PROJECT experiment \cite{UKDMC}, the
Micro-TPC Detector of the Kyoto-Tokyo collaboration \cite {KYOTOK}
and the TOKYO experiment \cite {TOKYO}.

The relative parameters $t$ and $h$ in the case
for light nuclear targets are essentially independent of the LSP mass, but they depend on the
energy cutoff, $Q_{min}$. For $Q_{min}=0$ they  are exhibited in
Table \ref{table1.gaus}. They are essentially the same for both the coherent and the spin modes.
 For intermediate and heavy nuclei they depend on the
LSP mass \cite{JDV03}.

In the case of the directional rates it is instructive to first
summarize our results regarding the reduction of the directional
rate compared to the usual rate, given by $\kappa/(2\pi)$. The
factors
 $\kappa$ depend, of course, on the angles of observation (see Table
 \ref{table1.gaus}).
 Second we should emphasize the importance of the modulation
 of the directional rates.
In the favored direction the modulation is not very large, but still it is
 three times larger compared to that of the non directional case.
 In the plane perpendicular to the sun's motion the modulation
is quite large (see Table \ref{table1.gaus}).

 Coming to transitions to excited states we believe that branching
 ratios of the size obtained here for $^{127}I$ are very
 encouraging to the experiments aiming at $\gamma$ ray detection,
 following the de-excitation of the nucleus.

 Regarding the detection of the emitted electrons in the LSP-nucleus
 collision we find  that, even though the distribution peaks at low energies,
 there remains
 substantial strength above $0.2~keV$, which is the threshold energy
 of a Micromegas detector, like the one recently
 \cite{GV03} proposed.

We thus  hope that, in spite of the  experimental difficulties,
some of  the above signatures  can be exploited by the
experimentalists.

\par
Acknowledgments: This work was supported in part by the European
Union under the contracts RTN No HPRN-CT-2000-00148 and TMR No.
ERBFMRX--CT96--0090.

\end{document}